\documentstyle[eqsecnum,preprint,aps,epsf]{revtex}

\begin{document}

\hoffset -1cm

\draft

\preprint{TPR-95-3}

\title{THE PLASMON IN HOT $\phi^4$ THEORY}

\author{Enke Wang$^*$ and Ulrich Heinz}

\address{
 Institut f\"ur Theoretische Physik, Universit\"at Regensburg,\\
  D-93040 Regensburg, Germany
}

\date{\today}

\maketitle

\begin{abstract}
  We study the 2-loop resummed propagator in hot $g^2\phi^4$ theory.
  The propagator has a cut along the whole real axis in the complex
  energy plane, but for small $g$, the spectral density is sharply
  peaked around the plasmon. The dispersion relation and the width of
  the plasmon are calculated at zero {\em and} finite momentum. At
  large momenta the spectral width vanishes, and the plasmon looses its
  collectivity and behaves like a non-interacting free particle.
\end{abstract}


\section{Introduction}
\label{s1}

At high temperature $T$ and low momenta $p < T$ the effective degrees
of freedom in any field theory are collective modes, bosonic
``plasmons" \cite{Wel,Re0} and fermionic ``plasminos''
\cite{Wel1,Rea}. Even if the fundamental field theoretical degrees of
freedom were massless, the collective modes possess a finite
``thermal" mass which is generated dynamically by the interactions
among the fundamental degrees of freedom. The thermal masses regulate
(some of) the severe infrared divergences which the massless theory
would otherwise develop at finite temperature due to the singularity
of the Bose distribution at zero momentum \cite{Kapusta,Reb}. The
strong rise of the Bose distribution at small momenta causes, for
example, the transport properties (viscosity, heat conduction) of a
hot plasma of massless fields to be dominated by the interaction
between low-momentum collective modes \cite{Re4,Re4c}. The
determination of their dispersion relation and their collisional width
is therefore a necessary prerequisite for any microscopic calculation
of the transport coefficients, for example using the Kubo formulae
\cite{Re3}.

It is known that a consistent determination of the plasmon dispersion
relation and width in a massless field theory requires the resummation
of certain loop contributions to the propagator, the so-called ``hard
thermal loops" (HTL) \cite{Reb}. In massless gauge theories, this
resummation becomes non-trivial beyond leading order, due to the
complicated analytical structure of the HTLs (i.e. their momentum
dependence and logarithmic branch cuts from Landau damping
\cite{Wel,Re0}). In massless scalar $\phi^4$ theory, the resummation
is much easier \cite{Dolan,Kapusta}, since the leading HTL (the
tadpole diagram) is just a momentum-independent real constant, and it
can therefore be relatively easily be carried beyond leading order
\cite{Re12}. Still, when we started this work, no calculation of the
collisional {\em width} of finite momentum plasmons existed. Since the
latter is needed, however, for the calculation of transport
coefficients from the Kubo formulae (previous calculations
\cite{Re4c,Re4b} used, for lack of better knowledge but without
justification, the zero momentum limit of the plasmon width), we
present in this paper a calculation of the resummed scalar field
propagator in $g^2\phi^4$ theory at the 2-loop level, including real
{\em and} imaginary parts in 4-dimensional momentum space. While this
work was in progress, Jeon \cite{Re4d} presented a calculation of the
plasmon width at zero and finite momentum on the plasmon mass shell.
Our results agree with his and extend them into the off-shell domain.

This paper is organized as follows. In Section \ref{s2} we will give a
brief review of results from Ref.~\cite{Re12} on resummation in hot
$\phi^4$ theory which we will need later. In Section \ref{s3} we
present the calculation of the imaginary part of the 2-loop
self-energy and investigate the analytical structure of the full
propagator at two-loop order. The properties of plasmons at rest and
at finite momentum are studied analytically and numerically in
Sections \ref{s3a} and \ref{s3b}, respectively. Our conclusions are
summarized in Section \ref{s4}. Appendix \ref{appa} contains some
technical steps of the calculations in Section \ref{s3}.

\section{Resummation at high temperature}
\label{s2}

For a hot scalar field, we consider the following Lagrangian with
$\lambda = {g^2 \over 4!} < 1$:
 \begin{equation}
 \label{lag0}
    {\cal L}_0 = {1\over 2} (\partial_{\mu}\phi)^2 +
                 {g^2 \over 4!} \phi^4 \, .
 \end{equation}
At the tree level it describes massless scalar fields with a quartic
self interaction. The induced thermal mass resulting from the single
HTL (tadpole) is of order $g$ and reads \cite{Dolan,Re12}
 \begin{equation}
 \label{mass}
   m_{\rm th}^2 = { g^2 T^2 \over 24} \, .
 \end{equation}
The effects from this thermal mass can be resummed by defining an
effective Lagrangian through
 \begin{equation}
 \label{lag}
    {\cal L} = \left( {\cal L}_0 + {1\over 2} m_{\rm th}^2
           \phi^2\right) - {1\over 2} m_{\rm th}^2 \phi^2
    = {\cal L}_{eff} - {1\over 2} m_{\rm th}^2 \phi^2\, ,
 \end{equation}
and treating the last term as an additional interaction. This
effective Lagrangian defines an effective propagator with thermal mass
$m_{\rm th}$
 \begin{equation}
 \label{delta}
   \Delta= {1 \over K^2+m_{\rm th}^2 } \, ,
 \end{equation}
where in the imaginary time formalism $K^2=k_0^2+{\bf k}^2$,
$k_0=2n\pi T$, $n=0,\pm1,\pm2,\dots$. It represents the infinite sum
of iterated HTLs. Since $m_{\rm th}$ leads to a qualitative
modification of the resummed effective propagator for soft momenta,
$p\le gT$, its inclusion via resummation is necessary to avoid
infrared divergences at finite temperature \cite{Kapusta}. Vertex
corrections from HTLs are down by a factor $g$ relative to the bare
vertex $g^2$ and don't qualitatively change the latter. In a
perturbative expansion of $\phi^4$ theory we can thus keep using the
bare vertex \cite{Re12}.

%
%

\begin{figure}
 \epsfxsize=10cm
 \centerline{\epsfbox{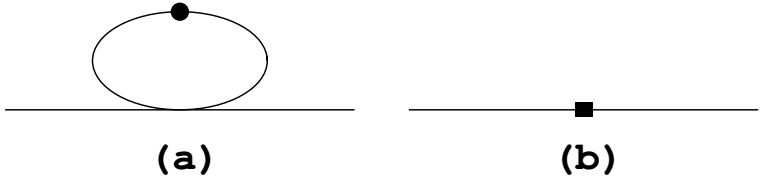}}
 \vskip 0.4cm
\caption{Lowest order contributions to the self energy for the
   resummed effective Lagrangian. Lines with a dot denote effective
   propagators with masses $m_{\rm th}$ or $m_{\rm p}$, respectively.
   The squared box denotes the additional interaction due to the mass
   shift.}
\label{F1}
\end{figure}

One can now recalculate the 1-loop self-energy using this resummed
effective propagator. This amounts to computing the contribution of
all ``cactus diagrams" or ``superdaisies" in the original Lagrangian
\cite{Dolan} each of which is infrared divergent. Using dimensional
regularization in the MS renormalization scheme, the diagram in
Fig.~\ref{F1}a gives
 \begin{equation}
 \label{sigma1p}
   \Sigma'_1 = {g^2 T^2 \over 24} \left( -1
       + { 3 m_{\rm th} \over \pi T }
          + { 3 \over 4\pi^2 }
            \left({ m_{\rm th} \over T } \right)^2
            \left[ \ln \left( { \mu^2 \over 4 \pi T^2 }\right)
            + \gamma_E \right] \right)
            + {\cal O}(g^5 \ln g)\, ,
 \end{equation}
where $\mu$ is the scale parameter in dimensional regularization and
$\gamma_E=0.577216...$ is Euler's constant. The resummation of an
infinite series of infrared diagrams reflects itself through a $g^3$
correction to the leading term $- g^2T^2/24 = - m_{\rm th}^2$; it is
non-analytic in the coupling constant $\lambda=g^2/24$ and for $g \ll
1$ dominates any finite genuinely perturbative corrections of order
$g^4$. In this sense resummation is essential.

Including also the contribution from the new 2-point interaction in
(\ref{lag}), Fig.~\ref{F1}b, the full 1-loop self-energy can be
written as $\Sigma_1=\Sigma'_1 + m_{\rm th}^2$. It is purely real.
Thus, to order $g^3$, the full propagator has only two poles at $p_0 =
\pm \sqrt{{\bf p}^2 + m_{\rm p}^2}$ where $m_{\rm p}$ is the 1-loop
resummed plasmon mass:
 \begin{equation}
 \label{mp}
    m_{\rm p}=m_{\rm th}\sqrt{1-{{3 m_{\rm th}}\over {\pi T}}}\, .
 \end{equation}

%
%

Following Ref.~\cite{Re12}, we can now improve the resummation
procedure by including into the effective Lagrangian all 1-loop
effects which are independent of the regularization scheme, i.e. by
writing
 \begin{equation}
 \label{lag1}
    {\cal L} = \left( {\cal L}_0 + {1\over 2} m_{\rm p}^2
           \phi^2\right) - {1\over 2} m_{\rm p}^2 \phi^2
 \end{equation}
and expanding into effective scalar propagators with an effective
plasmon mass $m_{\rm p}$. Then all factors $m_{\rm th}$ in
Eq.~(\ref{sigma1p}) are replaced by $m_{\rm p}$, and from the full
1-loop self energy $\Sigma_1 = \Sigma'_1 + m_{\rm p}^2$ all terms
up to order $g^3$ cancel, leaving a leading contribution of order
$g^4$ which is real.

\begin{figure}
 \epsfxsize10cm
 \centerline{\epsfbox{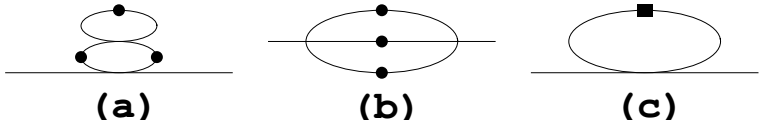}}
 \vskip 0.4cm
\caption{Next-to-lowest-order contributions to the self energy for the
   resummed effective Lagrangian.
}
\label{F2}
\end{figure}

At 2-loop order, the Feynman diagrams shown in Fig.~\ref{F2}
contribute. Their leading contributions are also of order $g^4$.
Diagram \ref{F2}c arises from the new 2-point interaction of the
Lagrangian (\ref{lag1}) and contributes at the same order.

The contribution $\Sigma'_2$ of the diagrams \ref{F2}a and \ref{F2}c
is easily evaluated. It is purely real. As a consequence of
resummation, all leading $g^4$ terms cancel, and the result is found
to be
 \begin{equation}
 \label{sigma2prime}
  \Sigma'_2 = {\cal O} (g^5\, \ln g)\, .
 \end{equation}
A convenient way to calculate diagram \ref{F2}b is to use the Saclay
method \cite{Re13}. Its contribution can be expressed as
 \begin{eqnarray}
 \label{sigma2pp}
   \Sigma''_2(ip_0,{\bf p}) &=& \int d[k,q] \,
        \Big\{ S(E_k,E_q,E_r)\lbrack
        (1+f_k)(1+f_q)(1+f_r)-f_kf_qf_r\rbrack
 \nonumber\\
   & &\phantom{\int d[k,q]}
      + S(E_k,E_q,-E_r)\lbrack(1+f_k)(1+f_q)f_r-f_kf_q(1+f_r)\rbrack
 \nonumber\\
   & &\phantom{\int d[k,q]}
      + S(E_k,-E_q,E_r)\lbrack(1+f_k)f_q(1+f_r)-f_k(1+f_q)f_r\rbrack
 \nonumber\\
   & &\phantom{\int d[k,q]}
      + S(-E_k,E_q,E_r)\lbrack f_k (1+f_q) (1+f_r)
                  -(1+f_k) f_q f_r \rbrack \Big\}  \, ,
 \end{eqnarray}
with the same notation as in Ref. \cite{Re12}, namely
 \begin{eqnarray}
 \label{dkq}
    d[k,q] &=& { g^4 \mu^{4\varepsilon} \over 6}\,
               { d^{D-1}k \over (2\pi)^{D-1} }\,
               { d^{D-1}q \over (2\pi)^{D-1} }\,
               { 1 \over 8 E_k E_q E_r } \, ,
 \\
    S(E_k,E_q,E_r) &=& {1 \over i p_0 + E_k + E_q + E_r }
                     - {1 \over i p_0 - E_k - E_q - E_r }  \, ,
 \end{eqnarray}
where
 \begin{eqnarray}
 \label{ref1}
   r &=& |{\bf k}+{\bf q}-{\bf p}| \, ,
 \\
 \label{ref2}
  E_l^2 &=& l^2 + m_{\rm p}^2\, , \qquad l=k,q,r,
 \\
 \label{ref3}
   f_l &=& { 1 \over \exp(E_l/T) - 1 }\, , \qquad l=k,q,r \, .
 \end{eqnarray}
$D=4-2\varepsilon$ is the the number of space-time dimension
introduced in dimensional regularization. Please notice that now the
one-loop resummed mass $m_{\rm p}$ occurs in all on-shell energies.
The real part of $\Sigma''_2$ contains an ultraviolet divergence. A
detailed discussion of the subtraction of this ultraviolet divergence
by a counter-term in the Lagrangian and the final expression for the
real part of the renormalized self-energy can be found in
Ref.~\cite{Re12}.  We can see from Eq.~(\ref{dkq}) that ${\rm Re\,}
\Sigma''_2$ is of the order $g^4$ and depends on the mass parameter
$\mu$ of the dimensional regularization.

The complete 2-loop expression for the real part of the self-energy
is given by
 \begin{equation}
 \label{sigma}
   {\rm Re\,} \Sigma(ip_0,{\bf p}) = \Sigma_1 + \Sigma'_2
            + {\rm Re\,} \Sigma''_2(ip_0,{\bf p}) \, ,
 \end{equation}
with the last term taken from Ref.~\cite{Re12}. A detailed calculation
of the imaginary part of the self-energy will be given in the
following section.

\section{Full propagator and spectral function of the plasmon}
\label{s3}

The leading contribution to the imaginary part of the self energy
comes from the 2-loop diagram \ref{F2}b. Making an analytical
continuation $ip_0=\omega+i\eta$, $\eta=0^+$, and using
$1/(A + i \eta) = {\cal P} (1/A) - i \pi \delta(A)$,
we obtain from Eq.~(\ref{sigma2pp})
 \begin{eqnarray}
 \label{imsigma}
    \lefteqn{ {\rm Im\,} \Sigma(\omega,{\bf p}) =
    {\rm Im\,} \Sigma''_2(\omega,{\bf p}) =-\pi\int d[k,q] }
 \\
    & &\Big\{\lbrack\delta(\omega+E_k+E_q+E_r)-\delta(\omega-E_k-E_q-
     E_r)\rbrack \lbrack(1+f_k)(1+f_q)(1+f_r)-
     f_kf_qf_r\rbrack
 \nonumber\\
    & &\! + \lbrack\delta(\omega+E_k+E_q-E_r)-\delta(\omega-E_k-
     E_q+E_r)\rbrack \lbrack(1+f_k)(1+f_q)f_r-
     f_kf_q(1+f_r)\rbrack
 \nonumber\\
    & &\! + \lbrack\delta(\omega+E_k-E_q+E_r)-\delta(\omega-E_k+E_q-
     E_r)\rbrack \lbrack(1+f_k)f_q(1+f_r)-f_k(1+f_q)f_r\rbrack
 \nonumber\\
    & &\! + \lbrack\delta(\omega-E_k+E_q+E_r)-\delta(\omega+E_k-E_q-
     E_r)\rbrack \lbrack f_k(1+f_q)(1+f_r)-(1+f_k)f_qf_r\rbrack\Big\}.
 \nonumber
 \end{eqnarray}
This gives the rates for physical decay and scattering processes
\cite{Re14}: The first line corresponds to the decay of one off-shell
into three on-shell $\phi$ mesons and (with opposite sign) the inverse
process; the two $\delta$-functions represent the decay of particles
with negative and positive energy, respectively. The remaining lines
in Eq.~(\ref{imsigma}) correspond to Landau damping via scattering of
the off-shell meson with on-shell particles from the heat bath,
forward and backward scattering process contributing with opposite
signs. Landau damping requires the presence of thermally excited
states and vanishes at zero temperature. In each term every ingoing
meson from the heat bath receives a thermal weight $f_i$ and each
outgoing one a Bose-enhancement factor $(1+f_i)$.

It is easy to show that
 \begin{equation}
 \label{resigma}
    {\rm Re\,} \Sigma(-\omega,{\bf p})
    = {\rm Re\,} \Sigma(\omega,{\bf p})
 \qquad {\rm and} \qquad
    {\rm Im\,} \Sigma(-\omega,{\bf p})
    = - {\rm Im\,} \Sigma(\omega,{\bf p}) \, ,
 \end{equation}
so we restrict ourselves to positive energy $\omega >0$.
After suitably relabeling the integration variables, the imaginary
part of the self-energy (\ref{imsigma}) for $\omega >0$ can be reduced
to
 \begin{equation}
 \label{img}
    {\rm Im\,}\Sigma(\omega,{\bf p}) = {\rm Im\,}g_1 (\omega,{\bf p})
                +{\rm Im\,}g_2 (\omega,{\bf p})
 \end{equation}
with the 3-body decay and Landau damping contributions, respectively:
 \begin{eqnarray}
 \label{img+1}
     {\rm Im\,} g_1(\omega,{\bf p})
     &=& \pi (e^{\omega/T}-1)
             \int d[k,q] f_k f_q f_r \delta(\omega-E_k-E_q-E_r)\, ,
 \\
 \label{img+2}
    {\rm Im\,} g_2(\omega,{\bf p})
    &=& 3 \pi (e^{\omega/T}-1)
            \int d[k,q] (1+f_k) f_q f_r \delta(\omega+E_k-E_q-E_r)\, .
 \end{eqnarray}
The $\delta$-functions in Eqs.~(\ref{img+1}) and (\ref{img+2}) arise
from energy conservation and constrain the integration over $k$ and
$q$ to the kinematically allowed domain. The evaluation of the
resulting kinematic limits of the integration variables is the most
demanding part of the following calculation and is shortly discussed
in Appendix \ref{appa}. Before going into the technical details of
evaluating the integrals, let us summarize some relevant analytical
properties of the propagators and self-energies. In the imaginary time
formalism, the full propagator
 \begin{equation}
 \label{d}
    D(ip_0,{\bf p}) = - {1 \over p_0^2 + {\bf p}^2 +
                         m_{\rm p}^2 - \Sigma(ip_0,{\bf p}) }
 \end{equation}
has the spectral representation
 \begin{equation}
 \label{ds}
    D(ip_0,{\bf p}) = \int_{-\infty}^{\infty} d\omega'\,
       { \rho(\omega',{\bf p}) \over  i p_0 - \omega' } \, .
 \end{equation}
The spectral density satisfies
 \begin{eqnarray}
 \label{rho}
    \rho(\omega,{\bf p}) &=& -{1\over {2\pi i}}
    \Big[ D(\omega+i\eta,{\bf p})-D(\omega-i\eta,{\bf p})\Big]
 \nonumber\\
    &=& {1\over \pi}\,{{{\rm Im\,}\Sigma(\omega,{\bf p})} \over
        {[\omega^2-{\bf p}^2-{m_{\rm p}}^2+
          {\rm Re}\Sigma(\omega,{\bf p})]^2 +
         [{\rm Im\,}\Sigma(\omega,{\bf p})]^2}} \, .
 \end{eqnarray}
It is an odd function of $\omega$, see Eqs.~(\ref{resigma}):
 \begin{equation}
 \label{rhoodd}
    \rho(-\omega,{\bf p}) = - \rho(\omega,{\bf p})
    \qquad {\rm or} \qquad
    \rho(\omega,{\bf p}) = \epsilon(\omega) \rho(|\omega|,{\bf p}) \, ,
 \end{equation}
where $\epsilon(\omega) = {\rm sgn}(\omega)$. Thus the full propagator
$D(ip_0,{\bf p})$ has the following analytical structure: If ${\rm
Im\, }\Sigma \ne 0$, the full propagator has a cut along the real axis
in the complex $\omega$ plane, with a discontinuity across this cut
given by the spectral function. If ${\rm Im\, }\Sigma$ goes to zero,
the identity
 \begin{equation}
 \label{da}
   {1\over \pi} \lim_{\eta \to 0} {\eta\over {A^2+\eta^2}}
    = \delta(A)
 \end{equation}
shows that the spectral density can be expressed as a
$\delta$-function at $\omega^2={\bf p}^2+m_{\rm p}^2-{\rm
Re\,}\Sigma(\omega, {\bf p})$; the corresponding analytical structure
of the full propagator $D(ip_0,{\bf p})$ are poles in the complex
$\omega$ plane given by the zeroes of the $\delta$-function.

{}From $D(ip_0,{\bf p})$ the full retarded propagator $D(\omega + i
\eta,{\bf p})$ is obtained by analytical continuation $i p_0 \to
\omega + i\eta$. The full propagator in the real time formalism can be
written as \cite{Re14}
 \begin{equation}
 \label{g}
   G(\omega, {\bf p})=D(\omega+i\epsilon(\omega)\eta)-
      2 \pi i {{\rho(|\omega|,{\bf p})}\over e^{|\omega|/T}-1 }\, .
 \end{equation}
All of these full propagators have the same analytical structure.

Defining the damping rate as
 \begin{equation}
 \label{gamma}
    \gamma(\omega,{\bf p}) =
    {{{\rm Im\,}\Sigma(\omega,{\bf p})}\over {2\omega}}
 \end{equation}
and substituting this into Eq.~(\ref{rho}), we see that if $\gamma(
\omega, {\bf p} ) \ll \sqrt{{\bf p}^2 +{m_{\rm p}}^2 -{\rm Re\,}
\Sigma(\omega,{\bf p})}$, then $\rho(\omega,{\bf p})$ is strongly
peaked at $\omega^2={\bf p}^2+{m_{\rm p}}^2-{\rm Re\,} \Sigma(\omega,
{\bf p})$. The solution of this dispersion relation defines a weakly
damped single-particle state which is interpreted as a quasiparticle,
the so-called plasmon. Denoting the solution for $\omega$ by
$\varepsilon({\bf p})$ and defining $\Gamma \equiv 2\gamma(
\varepsilon({\bf p}),{\bf p})$, the spectral density can in this limit
be expressed approximately in the form of a relativistic Breit-Wigner
function with full width $\Gamma$:
 \begin{equation}
 \label{rhobw}
    \rho(\omega,{\bf p}) \approx {1\over \pi}\,
    { \omega\Gamma \over
     (\omega^2 - \varepsilon^2({\bf p}))^2 + \omega^2 \Gamma^2 }
 \end{equation}
Finally, the spectral function fulfills the sum rule \cite{Re7}
 \begin{equation}
 \label{sumrule}
   2\int_0^{\infty}d\omega\, \omega\, \rho(\omega, {\bf p}) = 1\, .
 \end{equation}

In the following two subsections we now investigate in detail the
properties of plasmons in $\phi^4$ theory, both at rest and at finite
momentum ${\bf p}$ relative to the heat bath.

\subsection{Plasmons at rest, ${\bf p}=0$}
\label{s3a}

For ${\bf p}=0$ the evaluation of the imaginary part of the 2-loop
self-energy (\ref{img}) can be carried out in the following manner:
for the ${\bf q}$-integration we choose the $z$-axis along ${\bf k}$
and define $t=\cos({\bf k},{\bf q})$. The integrals in
Eqs.~(\ref{img+1}) and (\ref{img+2}) thus take the form
 \begin{equation}
 \label{int}
    \pi \int d[k,q] \longrightarrow  { g^4 \over 48(2\pi)^3 }
    \int_0^{\infty} dk\, k^2 \int_0^{\infty} dq\, q^2
    \int_{-1}^1 dt\, {1\over E_k E_q E_r}\, ,
 \end{equation}
where $E_r = \sqrt{k^2 + q^2 + 2 k q t + m^2_{\rm p}}$. The energy
conserving $\delta$-functions in Eqs.~(\ref{img+1},\ref{img+2}) will
be used to perform the $t$-integration by using
 \begin{equation}
 \label{delfunc}
   {{d E_r}\over {d t}}={{k q}\over E_r} \, .
 \end{equation}
The condition that the zeroes of the $\delta$-function fall into the
interval $-1 \leq t \leq 1$, i.e. $\sqrt{(k-q)^2+{m_{\rm p}}^2}\leq
E_r \leq \sqrt{(k+q)^2+{m_{\rm p}}^2}$, leads to restrictions of the
integration regions in $k$ and $q$. These restrictions are worked out
in Appendix \ref{appa}. For the 3-body decay contribution we find
 \begin{eqnarray}
 \label{img1}
   {\rm Im\,}g_1 (\omega,0)
   &=& \theta(\omega-3m_{\rm p}) \,
       {{g^4 T^2}\over 384 \pi^3 }\, \left( e^{\omega/T}-1 \right)
 \nonumber\\
   & &\times
      \int_a^{\varepsilon(x_1^*)} dv\, f(v) \,f(w-v)\,
      \ln\left( { f\left(\varepsilon(y_1)\right)
                  f\left(\varepsilon(y_2) + v - w \right)
                 \over
                  f\left(\varepsilon(y_2)\right)
                  f\left(\varepsilon(y_1) + v - w \right) } \right) \, ,
 \end{eqnarray}
where
 \begin{equation}
 \label{xya}
   a = { m_{\rm p} \over T}, \quad
   w = {\omega \over T}, \quad
   x = {k \over T}, \quad
   y = {q \over T}, \quad
   \varepsilon(z) = \sqrt{z^2+a^2}, \quad
   f(v) = {1 \over {e^v-1}}\, .
 \end{equation}
The specific values $x_1^*$ and $y_{1,2}$ are given by
Eqs.~(\ref{k1star}) and (\ref{qI},\ref{qII}) for ${\bf p}=0$,
respectively, divided $T$.

Similarly, the Landau damping contribution ${\rm Im\,}g_2 (\omega,0)$
is given by
 \begin{eqnarray}
 \label{imgg2}
   {\rm Im\,}g_2 (\omega,0)
   &=& {g^4 T^2 \over 128 \pi^3 }\, \left( e^{\omega/T}-1 \right)\,
       \left[ \theta(m_{\rm p}-\omega)
              \int_{\varepsilon(x_4^*)}^{\infty} dv \
            + \theta(\omega-m_{\rm p}) \int_a^{\infty} dv \right]
 \nonumber\\
   & &\times \Bigl(1+f(v)\Bigr) f(w+v) \,
     \ln\left( { f\left(\varepsilon(y_3)\right)
                 f\left(\varepsilon(y_4)-v-w\right)
               \over
                 f\left(\varepsilon(y_4)\right)
                 f\left(\varepsilon(y_3)-v-w\right) }\right) \, ,
 \end{eqnarray}
with $x^*_4$ and $y_{3,4}$ given by Eqs.~(\ref{k4star}) and
(\ref{qIII},\ref{qIV}) for ${\bf p}=0$, respectively, divided by $T$.

While ${\rm Im\,}g_1(\omega,0)$ vanishes below the 3-particle
threshold $E_0^*=3m_{\rm p}$ (see Eq.~(\ref{Epstar})), the Landau
damping contribution ${\rm Im\,}g_2(\omega,0)$ (which arises only at
$T\ne 0$) is nonzero for {\em all} positive values of $\omega$. Due to
the symmetry (\ref{resigma}), at 2-loop order the full propagator thus
has a cut along the whole real $\omega$ axis. This is qualitatively
different from the simple double pole structure at 1-loop order.

On the plasmon mass shell $\omega = m_{\rm p}$ only Landau damping
contributes to the imaginary part. From Appendix \ref{appa} we find
that in this case $B(m_{\rm p},k)=k^2$ and thus $y_3 = 0$, $y_4 = x$
as well as $x_4^* = 0$. Substituting this into Eq.~(\ref{imgg2}) and
using Eqs.~(\ref{img}) and (\ref{gamma}), we obtain the on-shell
plasmon damping rate (see also \cite{Re4d})
 \begin{equation}
 \label{gamma1}
    \gamma({m_{\rm p}},0)
    = { {\rm Im\,} \Sigma(m_{\rm p},0) \over
          2 m_{\rm p} }
    ={g^4 T^2 \over 256\pi^3 m_{\rm p}} L_2 (e^{-a})
    = { {g^3 T} \over {64 \sqrt{24} \pi} }\,
        \left[ 1 + {\cal O}(g\ln g) \right] \, ,
 \end{equation}
where $L_2(z)$ is the Spence function defined as
 \begin{equation}
 \label{spence}
   L_2(z) \equiv -\int_0^z dt\, {{\ln (1-t)}\over t}\, .
 \end{equation}
The leading term in the last expression of (\ref{gamma1}) was also
found in Ref.~\cite{Re12}.

${\rm Im\, } \Sigma$ is finite and does not require any
regularization. For dimensional grounds the width $\gamma(m_{\rm
p},0)$ for plasmons at rest is thus proportional to the temperature
$T$. From (\ref{gamma1}) we further see that for small $g$ the ratio
$\gamma(m_{\rm p},0)/T$ is proportional to $g^3$.

%
%

%
%

Using Eq.~(\ref{rho}) we can also evaluate the spectral function. We
notice that, as discussed in Section \ref{s2}, the ${\cal O}(g^4)$
term in ${\rm Re\,}\Sigma$ contains the renormalization scale $\mu$
which should be absorbed into the renormalized coupling constant. But
since the ${\cal O}(g^4)$ corrections have only a minimal effect on
the position of the peak in the spectral function, and all ${\cal
O}(g^3)$ terms have already been included in the plasmon mass, we can
to very good approximation neglect ${\rm Re\, } \Sigma$. In this
approximation the plasma frequency
 \begin{equation}
 \label{plfreq}
   \omega_{\rm p} = \varepsilon({\bf p}=0) = m_{\rm p}
   = {g T \over \sqrt{24}} \sqrt{ 1 - {3 g \over \pi
   \sqrt{24}}}
\end{equation}
is strictly proportional to $T$. An additional logarithmic temperature
dependence through the running of the coupling constant can only enter
in higher order.

\begin{figure}
 \epsfxsize10cm
\centerline{\epsfbox{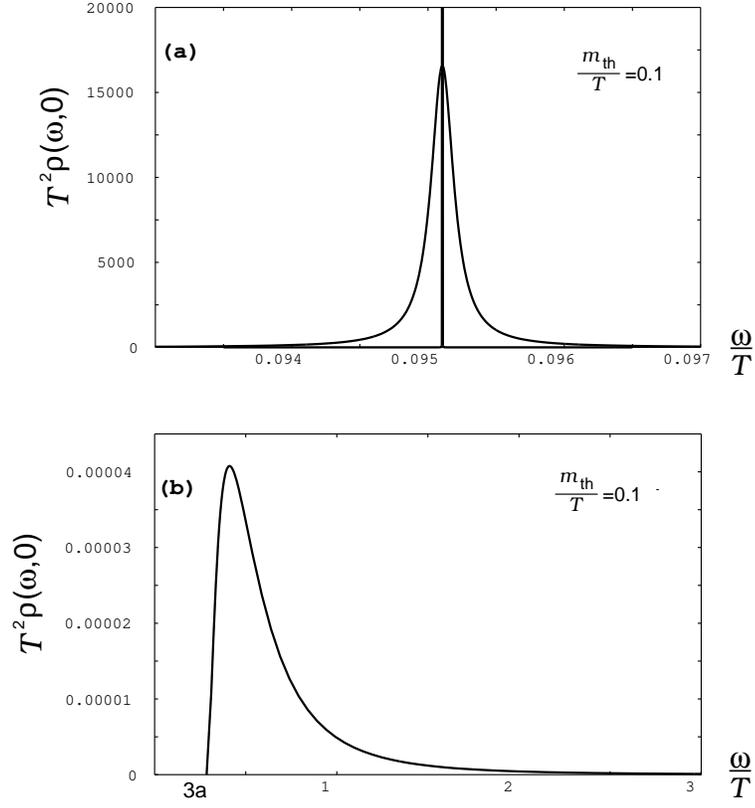}}
 \vskip 0.4cm
\caption{The two-loop spectral function for $m_{\rm th}/ T=0.1$. The
   vertical line at the peak position in the upper diagram indicates
   the spectral $\delta$-function at one-loop order.}
\label{F4}
\end{figure}

In the following we comment on some further properties of the plasmon
based on numerical results. In Fig.~\ref{F4}a we show for $m_{\rm th}/
T=0.1$ the 2-loop spectral function together with a line indicating
the pole position at one-loop order (including HTL resummation). Note
that there is no visible shift of the peak due to the higher order
corrections; the shift of the peak below the value of $m_{\rm th}$ is
entirely due to the 1-HTL-resummation, see Eq.~(\ref{plfreq}). The
major new feature at 2-loop order is just the finite width of the
plasmon. Although the 2-loop spectral density is strictly positive
for all values of $\omega$, it approaches zero very fast as one moves
away from the plasmon peak. Only in the neighborhood of the 3-body
decay threshold $\omega = 3\, m_{\rm p}$ it gains some weight again
as shown in Fig.~\ref{F4}b; the reader should notice, however, the
tiny vertical scale of this Figure.

\begin{figure}
 \epsfxsize10cm
\centerline{\epsfbox{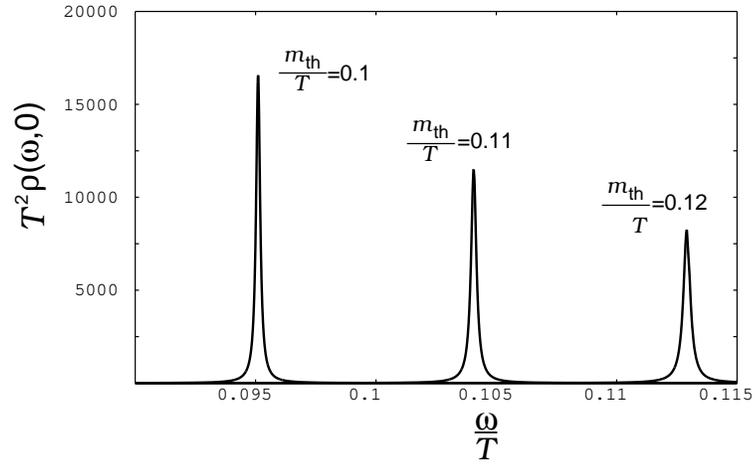}}
 \vskip 0.4cm
\caption{The two-loop spectral function for different values of $m_{\rm th}/T$
  (or coupling constant $g$).}
\label{F5}
\end{figure}

In Fig.~\ref{F5} we show the plasmon peak in the spectral function for
three different values of $m_{\rm th}/ T=0.10,\, 0.11,\, 0.12$. Since
both the plasmon frequency and its width grow linearly with $T$, and
according to Eq.~(\ref{rhobw}) $\rho(\omega_{\rm p}) \approx 1/(\pi
\omega_{\rm p} \Gamma)$, the height of the peak decreases with rising
temperature like $1/T^2$ for fixed $g$. This is seen in Fig.~\ref{F5}.
$T^2\rho(\omega_{\rm p})$ decreases accordingly with increasing
coupling constant like $1/g^4$ for small $g$.

For small $g$ the ratio of the damping rate to the plasma frequency is
 \begin{equation}
  {\gamma(m_{\rm p},0) \over \omega_{\rm p}}
  \approx {g^2 \over 64 \pi}\, .
 \label{ratio}
 \end{equation}
{}From Eq.~(\ref{plfreq}) we see that our resummation procedure is only
applicable for $\lambda = {g^2 \over 24} < {\pi^2 \over 9} \simeq 1$.
In this domain the ratio (\ref{ratio}) is always smaller than about
1/10, i.e. the plasmon is a well-defined quasiparticle which is not
overdamped.

\subsection{Moving Plasmons, ${\bf p}\ne 0$}
\label{s3b}

The evaluation of the imaginary part of the self-energy for ${\bf
p}\ne 0$ is a little more complicated because now two non-trivial
angular integrations occur. For the $q$-integration we choose the
$z$-axis in the direction of ${\bf k}-{\bf p}$ and define $t=\cos({\bf
k}-{\bf p},{\bf q})$. For the $k$-integration we choose the $z$-axis
along ${\bf p}$ and define $u=\cos({\bf p},{\bf k})$. In this way the
integrals in Eqs.~(\ref{img+1}) and (\ref{img+2}) can be expressed as
 \begin{equation}
 \label{intukqt}
    \pi\int d[k,q] \longrightarrow
    { g^4 \over 96 (2\pi)^3 } \int_0^{\infty} dk\, k^2 \int_{-1}^1 du
    \int_0^{\infty} dq \, q^2 \int_{-1}^1 dt\,
    {1\over E_k E_q E_r} \, .
 \end{equation}
The occurrence of two angles in the integrand renders the
determination of the integration limits from the energy-conserving
$\delta$-functions more complicated. However, as shown in Appendix
\ref{appa}, similar methods can be applied as in the case ${\bf p}=0$.
The $t$-integration in Eq.~(\ref{intukqt}) can be performed by using
 \begin{equation}
 \label{dert}
   {{d E_r}\over {d t}}={{q |{\bf k}-{\bf p}|}\over E_r} \, .
 \end{equation}
The condition that the zero of the energy $\delta$-function lies
inside integration domain leads to kinematic limits for the
integrations over $q$, $k$, and $u$. We find (see Appendix \ref{appa})
 \begin{eqnarray}
 \label{img1p}
     {\rm Im\,}g_1(\omega,{\bf p})
     &=& \theta\left( \omega - E_p^* \right)\,
         { g^4 T^2 \over 768 \pi^3}\, \left(e^{\omega/T}-1\right)
         \int_{-1}^1 du
         \int_0^{x_1^*} {dx\, x^2 \over |{\bf x} - {\bf z}|\,
                         \varepsilon(x)}
 \nonumber\\
     & &\times  f(\varepsilon(x))\, f(w-\varepsilon(x))\,
     \ln\left( { f(\varepsilon(y_1))\,
                 f(\varepsilon(y_2) + \varepsilon(x) - w )
               \over
                 f(\varepsilon(y_2))\,
                 f(\varepsilon(y_1) + \varepsilon(x) - w)}
        \right)
 \end{eqnarray}
with the same dimensionless variables (\ref{xya}) as before and
 \begin{equation}
 \label{z}
    {\bf z} = {{\bf p}\over T}\, .
 \end{equation}
The limits $x_1^*$, $y_1$, and $y_2$ are now given by
Eqs.~(\ref{kIstar}), (\ref{qI}), and (\ref{qII}), respectively,
divided by $T$. In the limit ${\bf p}\to 0$ ($z\to 0$)
Eq.~(\ref{img1p}) correctly reduces to Eq.~(\ref{img1}).

For the Landau damping contribution ${\rm Im\,}g_2(\omega,{\bf p})$ we
obtain
 \begin{eqnarray}
 \label{imgg2p}
   {\rm Im\,}g_2(\omega,{\bf p})
   &=& { g^4 T^2 \over 256 \pi^3 }\, \left(e^{\omega/T}-1\right)
       \int d(u,x) \, \bigl(1+f(\varepsilon(x))\bigr)\,
       f(w+\varepsilon(x))
 \nonumber\\
   & &\times \ln\left(
        { f(\varepsilon(y_3))\,
          f(\varepsilon(y_4)-\varepsilon(x)-w)
        \over
          f(\varepsilon(y_4))\,
          f(\varepsilon(y_3)-\varepsilon(x)-w) }
   \right) \, ,
 \end{eqnarray}
where
 \begin{eqnarray}
 \label{intku11}
 \lefteqn{ \int d(u,x) \equiv }
 \nonumber\\
  & & \phantom{ + }\theta(E_p - \omega)
        \Bigg[ \theta(\tilde E_p - \omega)
        \bigg( \theta(p - \omega) \int_{-\omega/p}^1 du
             + \theta(\omega - p) \int_{-1}^1 du        \bigg)
        \int_{x_4^*}^{\infty} dx
 \nonumber\\
  & & \phantom{+ \theta(E_p - \omega) }\!\!\!\!
     + \theta(\omega - \tilde E_p)
             \bigg( \int_{-1}^{u^*} du \int_{x_4^*}^{\infty} dx
           + \int_{u^*}^1 du \int_0^{\infty} dx \bigg)
      \Bigg] {x^2 \over |{\bf x}-{\bf z}|\, \varepsilon(x) }
 \nonumber\\
  & & +  \theta(\omega - E_p) {1\over z} \int_a^{\infty} d\varepsilon(x)
         \int_{|x-z|}^{|x+z|} d|{\bf x}-{\bf z}|
 \end{eqnarray}
if $p < 3 m_{\rm p}/2$ and
 \begin{eqnarray}
 \label{intku22}
 \lefteqn{ \int d(u,x) \equiv }
 \nonumber\\
  & & \phantom{ +} \theta(E_p - \omega) \Bigg[ \theta(\tilde E_p - \omega)
      \int_{-\omega/p}^1 du \int_{x_4^*}^\infty dx
 \nonumber\\
  & & \phantom{\theta(E_p - \omega) }
    + \theta(\omega - \tilde E_p) \Bigg( \theta(p - \omega)
              \bigg( \int_{-\omega/p}^{u^*} du \int_{x_4^*}^\infty dx +
              \int_{u^*}^1 du \int_0^\infty dx \bigg)
 \nonumber\\
  & & \phantom{\theta(E_p - \omega) \Bigg[ \theta(\tilde E_p - \omega)}
    + \theta(\omega - p) \bigg( \int_{-1}^{u^*} du
              \int_{x_4^*}^\infty dx + \int_{u^*}^1 du
              \int_0^\infty dx \bigg) \Bigg) \Bigg]
         {x^2 \over |{\bf x}-{\bf z}|\, \varepsilon(x) }
 \nonumber\\
  & & +  \theta(\omega - E_p) {1\over z} \int_a^{\infty} d\varepsilon(x)
         \int_{|x-z|}^{|x+z|} d|{\bf x}-{\bf z}|
 \end{eqnarray}
if $p > 3 m_{\rm p}/2$, respectively. In deriving these
expressions from the results (\ref{intku1}) and (\ref{intku2}) in
Appendix \ref{appa}, we used for the region $\omega > E_p$ the
relation
 \begin{equation}
 \label{dkpu}
   {{d |{\bf k}-{\bf p}|}\over {d u}}
   = -{{k \, p}\over |{\bf k}-{\bf p}|} \,
 \end{equation}
to rewrite the $u$-integration. Since the integrand in (\ref{imgg2p})
does not depend on $u$, the resulting integrals $\int d|{\bf x} - {\bf
z}|$ in (\ref{intku11}) and (\ref{intku22}) can be evaluated
trivially. Again ${\rm Im\,} g_2(\omega,{\bf p})$ is non-zero for all
$\omega >0$, and the full propagator has a cut along the whole real
$\omega$ axis, arising from Landau damping as for the case ${\bf
p}=0$. One easily checks that in the zero-momentum limit
Eq.~(\ref{imgg2p}) reduces to (\ref{imgg2}).

%
%

%
%

%
%

On the plasmon energy shell $\omega=\sqrt{{\bf p}^2+m_{\rm p}^2}$ the
plasmon damping rate can be deduced as
 \begin{eqnarray}
 \label{gammap}
    \gamma(\sqrt{{\bf p}^2+m_{\rm p}^2},{\bf p})
    &=& {{g^4 T}\over {256 \pi^3}}{1\over z\varepsilon(z)}
    \int_0^z dx \left[
      L_2(\xi)+L_2\left({{\xi-\zeta}\over {\xi (1-\zeta)}}\right)
      \right.
 \nonumber\\
    & & \left. - L_2\left({{\xi-\zeta}\over {1-\zeta}}\right)-
        L_2\left({{(\xi-\zeta)(1-\xi\zeta)}\over {\xi(1-\zeta)^2}}\right)
     \right] \, ,
 \end{eqnarray}
where
 \begin{equation}
 \label{xi}
    \xi=e^{-\sqrt{z^2+a^2}}\, ,\qquad  \zeta=e^{-\sqrt{x^2+a^2}} \, .
 \end{equation}
This on-shell result has also recently been obtained by
Jeon~\cite{Re4d} by a different method.

\begin{figure}
 \epsfxsize10cm
 \centerline{\epsfbox{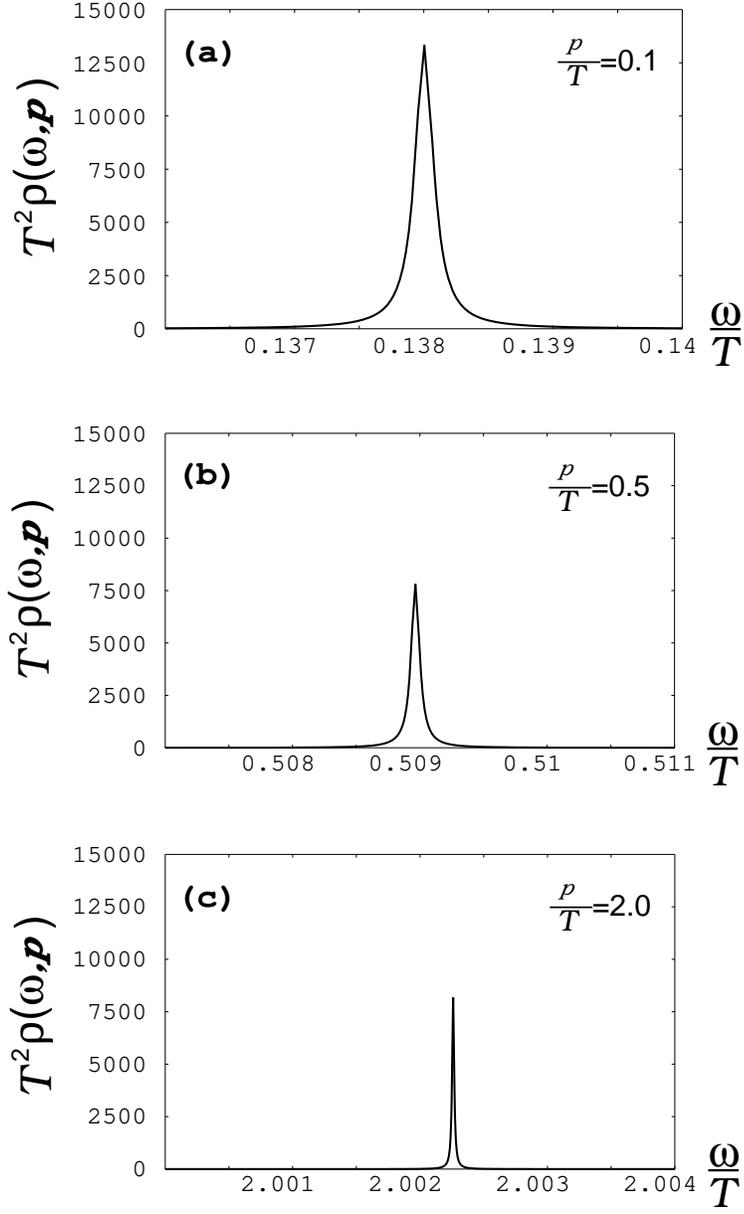}}
 \vskip 0.4cm
\caption{The two-loop spectral density at fixed $m_{\rm th} / T =0.1$ for
  different values of $p/T$.}
\label{F6}
\end{figure}

\begin{figure}
 \epsfxsize10cm
\centerline{\epsfbox{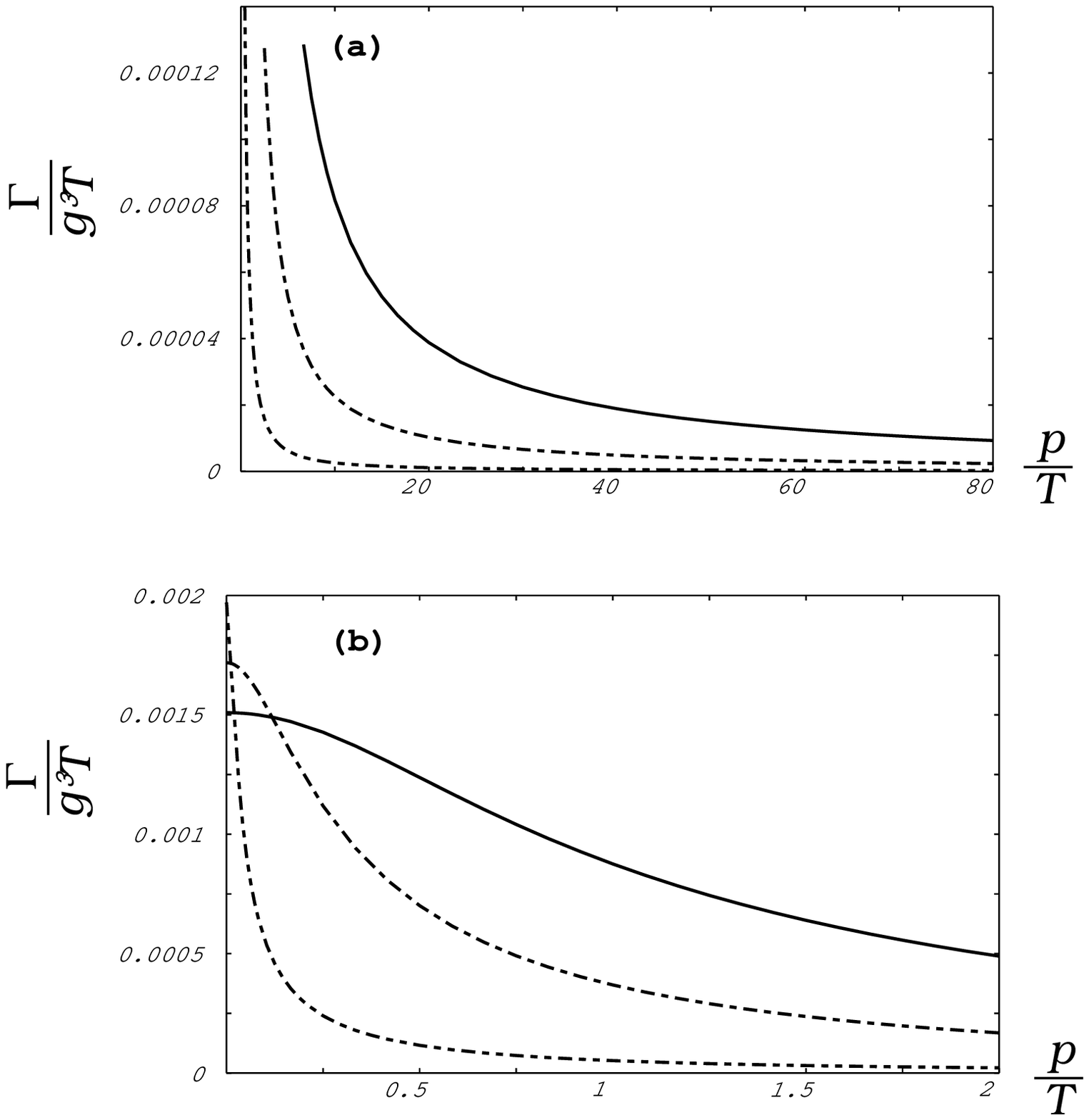}}
 \vskip 0.4cm
\caption{(a) The width of the spectral function vs. $p/T$ for
  $m_{\rm th}/T = 0.01$ (dot-dashed line), $m_{\rm th}/T =
  0.1$ (dashed line) and $m_{\rm th}/T = 0.5$ (solid line).
  (b) Close-up of (a) for small $p/T$.}
\label{F7}
\end{figure}

\begin{figure}
 \epsfxsize10cm
 \centerline{\epsfbox{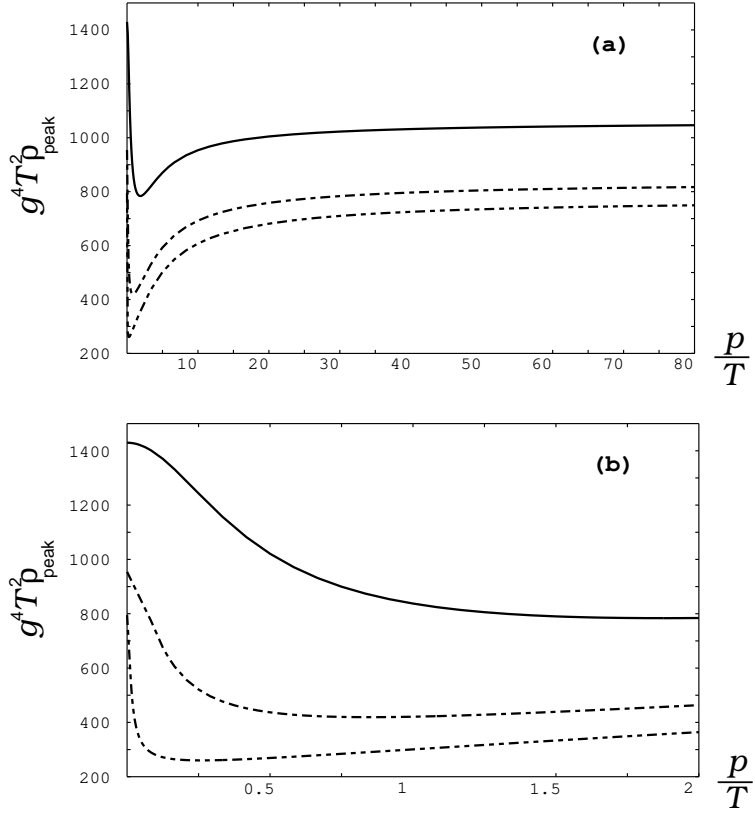}}
 \vskip 0.4cm
\caption{(a) The peak value of the spectral function vs. $p/T$ for
          $m_{\rm th}/T = 0.01$ (dot-dashed line) and $m_{\rm p}/T = 0.1$
          (dashed line) and $m_{\rm th}/T = 0.5$ (solid line).
   (b) Close-up of (a) for small $p/T$.}
\label{F8}
\end{figure}

For the numerical evaluation of the spectral function we substitute
Eqs.~(\ref{img1p}) and (\ref{imgg2p}) into Eq.~(\ref{img}) and then
use Eq.~(\ref{rho}). In Fig.~\ref{F6} we show the spectral density at
fixed $m_{\rm th}/T = 0.1$ for three values of the momentum $p/T$. As
$p/T$ increases, the width of the spectral function decreases as
illustrated in Fig.~\ref{F7}. For large $p/T$ the width goes to zero,
and at the same time the height of the spectral function approaches a
constant. This can been seen explicitly in Figs.~\ref{F7} and
\ref{F8}. This means that for large $p/T$ the plasmon looses the
features of a collective mode and again behaves like a non-interacting
free particle! At fixed, non-zero momentum ${\bf p}$ the dependences
on the temperature $T$ and the coupling constant $g$, on the other
hand, are qualitatively the same as at ${\bf p}=0$.

{}From the analysis of HTL resummation in gauge theories \cite{Reb} it
is known that there the summation scheme can only be guaranteed to
work in the soft momentum region $\omega,p \sim gT < T$. A partial
reason for this is that the thermal loops are momentum dependent, and
the particular momentum dependence of the ``hard" thermal loop arises
as the leading term in a high-temperature expansion of the 1-loop self
energy \cite{Wel}. Correspondingly, one might be inclined to distrust the
above results for momenta $p/T\gg 1$. In $\phi^4$ theory the situation
is, however, different in that the HTL which we resum is momentum
independent to order $g^2$ and $g^3$; momentum dependence enters only
at order $g^4$, and for weak coupling $\lambda = {g^2\over 24} \ll 1$
as in Figs. \ref{F6}, \ref{F7} it is weak on the scale $m_{\rm p}$.
Therefore the essential criterium for the validity of the resummation
procedure is the smallness of the coupling constant and not that of
$\omega/T$ or $p/T$.

%
%

This argument is also supported by a numerical investigation of the
sum rule Eq.~(\ref{sumrule}). In Table~\ref{T1} we give values for the
integral up to the 3-particle threshold,
 \begin{equation}
 \label{I}
   I_{\rm sum} = 2 \int_0^{E_p^*}
     d\omega\, \omega\, \rho(\omega, {\bf p})\, ,
 \end{equation}
for different values of $m_{\rm th} / T$ (or of the coupling constant
$\lambda=g^2/24$) at ${\bf p} = 0$ and for different values of $p/T$
at fixed $m_{\rm th} / T=0.1$. From the given numbers we see that in
all cases the sum rule is nearly exhausted by the collective plasmon
mode, and that the 3-body decay contribution at $\omega > E_p^*$ is
negligible. For large coupling constant ($m_{\rm th} / T > 1$), where
the resummation of hard thermal loops is no longer expected to be a
good approximation, we begin to see appreciable violations of the sum
rule. On the other hand, for $m_{\rm th}/T=0.1$, the sum rule is very
well satisfied even for very large momenta, $p/T \simeq 100$. As a
consequence, the shape of the spectral function (in particular the
interplay of the $p$-dependence of its width and height at the plasmon
pole shown in Figs.~\ref{F7} and \ref{F8}) is completely controlled by
this sum rule.

\section{Conclusions}
\label{s4}

In this paper we studied the analytical structure of the full
propagator and the properties of the plasmon in hot $\phi^4$ theory at
2-loop order, using the hard thermal loop (HTL) resummation scheme.

Since in $\phi^4$ theory the one-loop self-energy is completely real,
damping of the plasmon occurs only at the 2-loop level. Its physical
origin is Landau damping by scattering with thermal scalar field
quanta. This results in a cut of the full propagator along the entire
real frequency axis. However, for small coupling constant $g$ the
spectral function which is given by the discontinuity across this cut
shows a strong and rather sharp plasmon peak located at the position
expected from the 1-HTL-resummation. Its width $\Gamma/\omega_{\rm p}$
is approximately given by $g^2/(32\pi)$ for plasmons at rest in the
heat bath, and this width decreases even further if the plasmon is
moving relative to the heat bath. For very fast plasmons the width
goes to zero while the height of the plasmon peak approaches to a
constant: the plasmon behaves again like a non-interacting free
particle, i.e. it looses the features of a collective mode. Within the
range of validity of our resummation scheme, which is defined by the
smallness of the coupling constant $\lambda = {g^2\over 24}$ compared
to 1, the plasmon remains a well-defined, weakly-damped quasiparticle
excitation for all values of the temperature. The sum rule
Eq.~(\ref{sumrule}) for the spectral density is very well satisfied by
the 2-loop result even for momenta much larger than $T$, and it is
essentially saturated by the plasmon peak.

\acknowledgments
We are grateful to U. Wiedemann for helpful discussions, and we would
like to thank the referee for a careful reading of the manuscript and
for pointing out an error in the original version of the paper. This
work was supported in part by the Deutsche Forschungsgemeinschaft
(DFG), the Bundesministerium f\"ur Bildung und Forschung (BMBF), the
National Natural Science Foundation of China (NSFC) and the
Gesellschaft f\"ur Schwerionenforschung (GSI).


\appendix
\section{Kinematic limits for the collision integrals}
\label{appa}

Here we give some technical details on how to solve for the kinematic
limits of the integral (\ref{intukqt}) resulting from the energy
conserving $\delta$-functions. We perform the calculation for
arbitrary external momentum ${\bf p}$; the expressions needed in
Section \ref{s3a} for plasmons at rest are easily obtained by setting
${\bf p}=0$. In this limit the integrand is independent of $u$, and
the $u$-integration gives a trivial factor of 2.

We start by doing the $t$-integration in (\ref{intukqt}) using
(\ref{dert}) and exploiting the energy $\delta$-function. In order to
obtain a non-zero contribution, the solution $E_r$ of the
$\delta$-function constraint must lie in the interval
 \begin{equation}
 \label{Eint}
   \sqrt{(|{\bf k}-{\bf p}| - q)^2+{m_{\rm p}}^2}
   \leq E_r \leq
   \sqrt{(|{\bf k}-{\bf p}| + q)^2+{m_{\rm p}}^2}
 \end{equation}
defined by the limits of the $t$ integration, $-1 \leq t \leq 1$. For
fixed $\vert {\bf k-p} \vert$ (or fixed $k$ and $u$) these limits on
$E_r$ generate upper and lower integration limits for $q$. If the
latter are complex, the corresponding values of $k$ and $u$ must be
excluded from the exterior integrations.

\begin{figure}
 \epsfxsize10cm
\centerline{\epsfbox{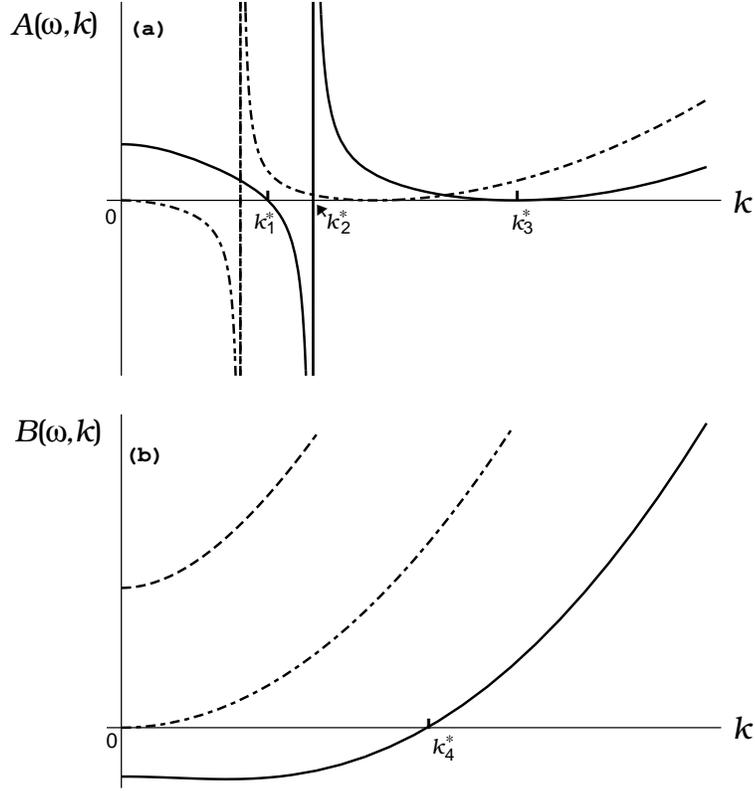}}
 \vskip 0.4cm
\caption{(a) The function $A(\omega,k)$ vs. $k$ for $\omega > 3 m_{\rm p}$
   (solid line) and for $\omega =3m_{\rm p}$ (dot-dashed line);
   $k_1^*$ and $k_3^*$ indicate the two zeroes and $k_2^*$ the
   singularity of the solid curve. (b) The function $B(\omega,k)$ vs.
   $k$ for $\omega <m_{\rm p}$ (solid line), for $\omega =m_{\rm p}$
   (dot-dashed line) and for $\omega >m_{\rm p}$ (dashed line);
   $k_4^*$ indicates the zero of the solid curve.}
\label{F3}
\end{figure}

For the 3-body decay (\ref{img+1}) we must solve $\omega-E_k-E_q-E_r =
0$, with the notation of Eqs.~(\ref{ref1}), (\ref{ref2}). This has no
solutions unless $\omega$ exceeds the 3-particle threshold
 \begin{equation}
 \label{Epstar}
   E_p^* = m_{\rm p} + \sqrt{p^2 + 4 m_{\rm p}^2}  \, .
 \end{equation}
With the help of the auxiliary function
 \begin{equation}
 \label{funcAA}
    A(\omega,|{\bf k}-{\bf p}|) = { (\omega - E_k)^2\,
     \bigl((\omega - E_k)^2 - ({\bf k}-{\bf p})^2 - 4 m_{\rm p}^2 \bigr)
                     \over
                    (\omega - E_k)^2 - ({\bf k}-{\bf p})^2 }
 \end{equation}
the lower and upper integration limits for $q$ which result from
(\ref{Eint}) can be expressed as
 \begin{equation}
 \label{qI}
   q_1 = \max [0,q'_1]  \qquad {\rm with} \qquad
   q'_1 = {1\over 2} \left( \sqrt{ A(\omega, |{\bf k}-{\bf p}|)}
           - |{\bf k}-{\bf p}| \right)
 \end{equation}
and
 \begin{equation}
 \label{qII}
    q_2 = {1\over 2} \left( \sqrt{ A(\omega,|{\bf k}-{\bf p}|)}
          + |{\bf k}-{\bf p}| \right) \, .
 \end{equation}
For $q_1$, $q_2$ to be real, $A(\omega,|{\bf k}-{\bf p}|)$ must be
positive. The function $A$ is shown in Fig.~\ref{F3}a. $A(\omega,|{\bf
k}-{\bf p}|)$ has a single zero at
 \begin{equation}
 \label{kIstar}
    k_1^* = { pu (\omega^2 - p^2 - 3\, m_{\rm p}^2)
              + \omega \sqrt{ (\omega^2 - p^2 - 3\, m_{\rm p}^2)^2
              - 4(\omega^2 - p^2 u^2) m_{\rm p}^2}
              \over 2 (\omega^2 - p^2 u^2) } \, ,
 \end{equation}
which for ${\bf p}=0$ reduces to
 \begin{equation}
 \label{k1star}
    k_1^* = { \sqrt{(\omega^2-m_{\rm p}^2)
                    (\omega^2 - 9 m_{\rm p}^2)} \over
              2 \omega } \, ,
 \end{equation}
a ${\bf p}$-independent double zero at
 \begin{equation}
 \label{k3star}
    k_3^* = \sqrt{\omega^2-m_{\rm p}^2} \, ,
 \end{equation}
and a singularity in between at
 \begin{equation}
 \label{kIIstar}
   k_2^* = { pu (\omega^2 - p^2 + m_{\rm p}^2)
             + \omega \sqrt{ (\omega^2 - p^2 + m_{\rm p}^2)^2
              - 4(\omega^2 - p^2 u^2) m_{\rm p}^2}
             \over  2 (\omega^2 - p^2 u^2) } \, ,
 \end{equation}
which for ${\bf p}=0$ becomes simply
 \begin{equation}
 \label{k2star}
    k_2^* = { \omega^2 - m_{\rm p}^2 \over 2 \omega } \, .
 \end{equation}
One checks that for $k > k_2^*$ the energy constraint cannot be
satisfied; for $k_1^* < k < k_2^*$ the function $A(\omega,|{\bf k}-
{\bf p}|)$ is negative, giving rise to complex $q$ integration limits.
Hence the allowed region for $k$ is $0 \leq k \leq k_1^*$. There is no
constraint on $u$. The $q$-integration can be done analytically,
giving the results presented in Section \ref{s3}.

%
%

For the Landau damping term (\ref{img+2}) the calculation is slightly
more involved. We must solve $\omega-E_k-E_q-E_r = 0$ with the
restriction (\ref{Eint}). For fixed $\vert {\bf k-p}\vert$ the
resulting lower and upper limits for $q$ can now be expressed in terms
of another auxiliary function
 \begin{equation}
 \label{funcBB}
    B(\omega,|{\bf k}-{\bf p}|) = { (\omega + E_k)^2\,
     \bigl((\omega + E_k)^2 - ({\bf k}-{\bf p})^2 - 4 m_{\rm p}^2 \bigr)
                     \over
                    (\omega + E_k)^2 - ({\bf k}-{\bf p})^2 }
 \end{equation}
as
 \begin{equation}
 \label{qIII}
    q_3 = \max [0,q'_3] \qquad {\rm with} \qquad
    q'_3 = {1\over 2} \left( \sqrt{ B(\omega, |{\bf k}-{\bf p}|) }
           - |{\bf k}-{\bf p}| \right) \, ,
 \end{equation}
and
 \begin{equation}
 \label{qIV}
    q_4 = {1\over 2} \left( \sqrt{ B(\omega,|{\bf k}-{\bf p}|) }
        + |{\bf k}-{\bf p}| \right) \, .
 \end{equation}
The function $B(\omega,k)$ is illustrated in Fig.~\ref{F3}b.
For $\omega \geq E_p$, $B(\omega, |{\bf k}-{\bf p}|)$ is
always positive, so $q_3, q_4$ are real and the $\delta$-function can
be satisfied for all $k\in [0,\infty]$ and $u\in [-1,1]$. For $\omega
< E_p$, $B(\omega,|{\bf k}-{\bf p}|)$ has a positive and real zero at
 \begin{equation}
 \label{kIIIstar}
   k_4^* = { pu (\omega^2 - p^2 - 3 m_{\rm p}^2)
             + \omega \sqrt{ (\omega^2 - p^2 - 3 m_{\rm p}^2)^2
             - 4(\omega^2 - p^2 u^2) m_{\rm p}^2 }
             \over 2 (\omega^2 - p^2 u^2) } \, .
 \end{equation}
if simultaneously $u < u^*$ with
 \begin{equation}
 \label{ustar}
   u^* = - { \sqrt{ 4\, \omega^2\, m_{\rm p}^2
                  - (\omega^2 - p^2 - 3 m_{\rm p}^2)^2 }
             \over  2\, p\, m_{\rm p} } \, .
 \end{equation}
$u^*$ itself is real only for $\omega \geq \tilde E_p$ where
 \begin{equation}
 \label{Ep}
   \tilde E_p = \sqrt{p^2 + 4 m_{\rm p}^2} - m_{\rm p}\, .
 \end{equation}
For $\tilde E_p < \omega < E_p$, positivity of $B(\omega, |{\bf k}-
{\bf p}|)$ thus restricts the integration range of $k$ to $k > k_4^*$
if $u < u^*$, while all $k$ are allowed for $u > u^*$. For $\omega <
\tilde E_p$ there is no upper constraint on $u$, and $k$ is restricted
to the region $k > k_4^*$.

For ${\bf p}=0$, there is no constraint on $u$, and $k_4^*$ reduces to
 \begin{equation}
 \label{k4star}
   k_4^* = { \sqrt{ (m_{\rm p}^2-\omega^2)
                    (9 m_{\rm p}^2 - \omega^2)} \over
             2 \omega } \, .
 \end{equation}

We can summarize the results so far by writing
 \begin{eqnarray}
 \label{intku}
    \int du \int dk
    & \longrightarrow &
      \theta(E_p - \omega) \Bigg[ \theta(\tilde E_p - \omega)
      \int^1 \!\! du \int_{k_4^*}^{\infty} dk
 \nonumber\\
    & & \phantom{\theta(E_p - \omega) }\!\!
      + \theta(\omega-\tilde E_p)
               \left( \int^{u^*} \!\! du \int_{k_4^*}^{\infty} dk
                    + \int_{u^*}^1 du \int_0^{\infty} dk \right) \Bigg]
 \nonumber\\
    & & \!\! + \theta(\omega - E_p) \int_{-1}^1 du \int_0^{\infty} dk \, .
 \end{eqnarray}
The missing {\em lower} limits of the $u$-integration in two of the
terms will be determined now.

Eq.~(\ref{kIIIstar}) shows that for $\omega < p$, $k_4^*$ has a
singularity at $u= - \omega/p$. (At $u= +\omega/p$, $k_4^*$ remains
finite because the numerator vanishes, too.) This implies an
additional restriction on the lower limit of the $u$ integration
domain for the two $k_4^*$-dependent terms in Eq.~(\ref{intku})
whenever $\omega < p$. Which of these two terms is affected depends on
the magnitude of $p$: For $p < 3 m_{\rm p}/2$ we find also $p < \tilde
E_p$, so in this case the second term in Eq.~(\ref{intku}) is not
affected by this problem because it receives contributions only from
$\omega > \tilde E_p > p$. For $p > 3 m_{\rm p}/2$ we get also $p >
\tilde E_p$, and then both the first and the second term in
Eq.~(\ref{intku}) are affected. It turns out that the region $u <
-\omega/p$ is always unphysical because either the function $B$ is
negative or the $\delta$-function constraint cannot be satisfied. One
thus finds, using the dimensionless quantities (\ref{xya}),
 \begin{eqnarray}
 \label{intku1}
 \lefteqn{ \int du \int dk
           \stackrel{^{p<3 m_{\rm p}/2}}{\longrightarrow} }
 \nonumber\\
  & & \phantom{ + }\theta(E_p - \omega)
        \Bigg[ \theta(\tilde E_p - \omega)
        \bigg( \theta(p - \omega) \int_{-\omega/p}^1 du
             + \theta(\omega - p) \int_{-1}^1 du        \bigg)
        \int_{x_4^*}^{\infty} dx
 \nonumber\\
  & & \phantom{+ \theta(E_p - \omega) }\!\!\!\!
     + \theta(\omega - \tilde E_p)
             \bigg( \int_{-1}^{u^*} du \int_{x_4^*}^{\infty} dx
           + \int_{u^*}^1 du \int_0^{\infty} dx \bigg)
      \Bigg]
 \nonumber\\
  & & +  \theta(\omega - E_p) \int_{-1}^1 du \int_0^{\infty} dx
 \end{eqnarray}
and
 \begin{eqnarray}
 \label{intku2}
 \lefteqn{ \int du \int dk
           \stackrel{^{p>3 m_{\rm p}/2}}{\longrightarrow} }
 \nonumber\\
  & & \phantom{ +} \theta(E_p - \omega) \Bigg[ \theta(\tilde E_p - \omega)
      \int_{-\omega/p}^1 du \int_{x_4^*}^\infty dx
 \nonumber\\
  & & \phantom{\theta(E_p - \omega) }
    + \theta(\omega - \tilde E_p) \Bigg( \theta(p - \omega)
              \bigg( \int_{-\omega/p}^{u^*} du \int_{x_4^*}^\infty dx +
              \int_{u^*}^1 du \int_0^\infty dx \bigg)
 \nonumber\\
  & & \phantom{\theta(E_p - \omega) \Bigg[ \theta(\tilde E_p - \omega)}
    + \theta(\omega - p) \bigg( \int_{-1}^{u^*} du
              \int_{x_4^*}^\infty dx + \int_{u^*}^1 du
              \int_0^\infty dx \bigg) \Bigg) \Bigg]
 \nonumber\\
  & & + \theta(\omega - E_p) \int_{-1}^1 du \int_0^{\infty} dx
  \, .
 \end{eqnarray}



\newpage

\newpage

\begin{table}
\caption{The values of $I_{\rm sum}$ for different values of
         $m_{\rm th}/T$ at fixed ${\bf p}{=}0$ (upper half) and for
         different values of $p/T$ at fixed $m_{\rm th}/T=0.1$ (lower half).
         }
\begin{tabular}{lcccc}
$m_{\rm th}/T$ & 0.05 & 0.1 & 0.5 & 1.0\\
$I_{\rm sum}$ & 0.9834 & 0.9852 & 0.9916 & 1.0669 \\
\tableline
$p/T$ & 0. & 10. & 20. & 80.\\
$I_{\rm sum}$ & 0.9852 & 0.9896 & 0.9903 & 0.9914 \\
\end{tabular}
\label{T1}
\end{table}


\end{document}